\pdfoutput=1
\documentclass[11pt]{article}

\usepackage[T1]{fontenc}
\usepackage[utf8]{inputenc}
\usepackage{lmodern}
\usepackage{cite}
\usepackage{amsmath,amssymb,amsfonts}
\usepackage{graphicx}
\usepackage{textcomp}
\usepackage{booktabs}
\usepackage{array,tabularx}
\usepackage{ragged2e}
\usepackage{bm}
\usepackage{authblk}
\usepackage[a4paper,margin=1in]{geometry}
\usepackage{microtype}
\usepackage[hidelinks]{hyperref}

\def\BibTeX{{\rm B\kern-.05em{\sc i\kern-.025em b}\kern-.08em
    T\kern-.1667em\lower.7ex\hbox{E}\kern-.125emX}}

\title{Computation-Limited Signals in Practice: A Throughput Anomaly in Software-Defined O-RAN PHY Layers}

\author[1]{Gabriel Alessi Posonski}
\author[1,3]{Saulo Queiroz}
\author[2,3]{João P. Vilela}
\author[4]{Benjamin Koon Kei Ng}
\author[4]{Chan-Tong Lam}
\author[3]{Edmundo Monteiro}

\affil[1]{Academic Department of Informatics, Federal University of Technology (UTFPR), Ponta Grossa, PR, Brazil}
\affil[2]{CRACS/INESC TEC and Department of Computer Science, Faculty of Sciences, University of Porto, 4169-007 Porto, Portugal}
\affil[3]{Centre for Informatics and Systems of the University of Coimbra (CISUC), Coimbra, Portugal}
\affil[4]{Faculty of Applied Sciences, Macao Polytechnic University, Macao}
\affil[]{\textit{E-mail:} \texttt{gabrielalessi@alunos.utfpr.edu.br}, \texttt{sauloqueiroz@utfpr.edu.br}}
\affil[]{\texttt{jvilela@fc.up.pt}, \texttt{\{saulo,jpvilela,edmundo\}@dei.uc.pt}}
\affil[]{\texttt{\{bng,ctlam\}@mpu.edu.mo}}
\affil[]{Gabriel Alessi Posonski and Saulo Queiroz contributed equally to this work.}
\affil[]{Corresponding author: Saulo Queiroz (\href{mailto:sauloqueiroz@utfpr.edu.br}{sauloqueiroz@utfpr.edu.br}).}
\date{}

\begin{document}
\maketitle

\begin{abstract}
In this work, we formalize a performance anomaly in software-based wireless physical-layer (PHY) implementations, a key enabling technology for Open Radio Access Network (O-RAN) systems. For a fixed offered load, the PHY burst length--that is, the number of PHY data units, or symbols, required to transmit the load--decreases as the 
nominal PHY bit rate increases because each unit conveys more bits. Shorter bursts, however, provide fewer opportunities to exploit the concurrent operation of pipelined modulation, transmission, and demodulation stages. Consequently, processing latency accounts for an increasingly large fraction of the total burst delivery time and may cause a nominally faster signal PHY bit rate to deliver lower effective throughput. We refer to the operating condition in which computational latency governs the effective throughput as the computation-limited (comp-limited) regime. The most pronounced manifestation of this regime occurs when the selected PHY bit rate allows the entire offered load to be transmitted in a single symbol. In this limiting case, intersymbol pipelining cannot be exploited, and the modulation and demodulation latencies cannot be amortized over multiple symbols. Consequently, their combined processing latency may exceed the symbol duration even when the transmitter and receiver independently satisfy their real-time processing requirements. Moreover, because the computational complexity of PHY signal processing often scales superlinearly with the resources allocated to each symbol, a configuration with a nominally higher PHY bit rate may deliver lower throughput than a nominally slower configuration due to the greater processing-latency overhead associated with the higher-rate configuration. We validate this anomaly using processing-latency measurements reported in the literature for established software-defined PHY implementations. Our results indicate that PHY processing latency should be explicitly incorporated into link-adaptation algorithms for very-high-throughput services in future 6G wireless networks.
\end{abstract}

\medskip

\section{Introduction}
\label{sec:introduction}
Wireless communication systems continue to pursue higher data rates 
through wider bandwidths, denser modulation, more spatial streams, and increasingly
 sophisticated coding and detection algorithms \cite{saad2019vision}. These mechanisms
 improve the nominal physical-layer (PHY) bit rate, but they also increase the computational
complexity to process each PHY data unit (or symbol), generally, at an asymptotically
higher pace in comparison to the symbol's resources. 
For example, increasing the number of subcarriers \(N\) in an orthogonal frequency-division 
multiplexing (OFDM) symbol allows more bits to be conveyed per symbol but increases the computational 
complexity of the fast Fourier transform (FFT) superlinearly, as \(O(N\log_2 N)\). Similar `side-effect' 
performance arises in other computationally intensive PHY operations, such as channel estimation and channel 
decoding, depending on the adopted algorithms and signal configurations. 

To sustain real-time point-to-point PHY operation on a per-symbol basis, a PHY implementation must 
complete the computational processing associated with each symbol within its prescribed deadline, which must 
be no longer than one symbol interval. Consider, for example, the \(4~\mu\mathrm{s}\) symbol duration of the 
legacy IEEE 802.11g OFDM PHY. To sustain real-time operation on a per-symbol basis, both modulation and demodulation 
must each be completed within \(4~\mu\mathrm{s}\). Therefore, the end-to-end latency required to modulate, transmit, 
and demodulate the first symbol can be as large as three symbol durations, corresponding to \(12~\mu\mathrm{s}\)
 in this example. 

In practice, however, transmitting a given offered load typically requires multiple symbols. These symbols 
form a burst and traverse the modulation, transmission, and demodulation stages in a pipelined fashion. Consequently, 
modulation and demodulation latencies are amortized over multiple symbols and are therefore neglected by the conventional 
bit-rate model~\cite{proakis2008digital}. This abstraction, however, does not accurately characterize the end-to-end 
delivery time of the first symbol. To address this limitation, the authors of~\cite{queiroz2021,queiroz2024,itu} introduced the 
spectro-computational (SC) analysis, a mathematical framework that incorporates PHY processing latency arising from 
computational complexity into conventional communication-performance metrics.

The SC framework can assist PHY designers in balancing the tradeoff between computational 
complexity and communication performance when developing new waveforms and signal-processing algorithms. 
From this perspective, previous studies have investigated OFDM with index modulation~\cite{index-selector7,optimal-mapper9,maximal-se} 
and the computational scalability of fast Fourier transform implementations~\cite{fast-enough11,queiroz-ieeespmagazine-2025}.

Moreover, the SC framework formalizes an asymptotic communication regime in which the intrinsic lower bound 
on the computational complexity required to process the symbol of a given waveform grows asymptotically faster than 
the number of bits modulated onto that symbol. Under these conditions, the baseband processing capacity must scale accordingly;
 otherwise, the effective PHY throughput approaches zero as increasingly large channel resources
are allocated to the symbol. This regime is referred to as computation-limited (or comp-limited), reflecting the 
fact that system performance ultimately remains constrained by computational resources even when conventional 
channel resources, such as bandwidth and transmit power, become arbitrarily abundant.

In this work, we identify a practical manifestation of the comp-limited regime in the context of Open Radio 
Access Network (O-RAN) systems~\cite{o-ran}. Software-based PHY implementations play a key enabling role in O-RAN 
by allowing network functionality to be upgraded while minimizing the need to replace the underlying hardware infrastructure.

Our key finding concerns scenarios in which a fixed offered load can be conveyed using only a 
few PHY symbols--or even a single symbol--a condition that becomes increasingly likely at high nominal bit rates. 
As the nominal bit rate increases, the number of symbols per burst decreases because each symbol conveys more bits. 
Consequently, processing latency is amortized over fewer symbols and accounts for an increasingly large fraction of 
the total burst delivery time. This effect is compounded by the greater processing cost of higher-rate signal configurations, 
particularly when the computational complexity of the underlying PHY algorithms scales superlinearly with relevant signal dimensions,
as previously explained.

Focusing on the limiting case of a single-symbol burst, we analyze published processing-latency measurements 
from established real-time software-based PHY implementations and demonstrate a throughput anomaly whereby a nominally 
higher-rate configuration delivers lower effective PHY throughput than a nominally slower configuration, reversing the 
performance ranking implied by conventional bit-rate metrics. 

In summary, the main contributions of this work are as follows:
\begin{itemize}
\item an extension of the SC throughput model that explicitly parameterizes the burst length;
\item a numerical evaluation of multiple bit-rate configurations using a well-established legacy Wi-Fi PHY implementation~\cite{sora};
\item the derivation of an explicit criterion for the throughput anomaly and the identification of the payload-size intervals over which a lower nominal bit rate yields higher effective throughput.
\end{itemize}

The remainder of this article is organized as follows. Section~\ref{sec:background} reviews SC analysis and software-implemented PHYs. Section~\ref{sec:model} derives the expanded throughput model. Section~\ref{sec:method} describes the implementation survey, data-selection rationale, and processing-time calculation. Section~\ref{sec:results} presents the numerical results. Section~\ref{sec:discussion} discusses implications and limitations, and Section~\ref{sec:conclusion} concludes the article.

\section{Background and Related Work}
\label{sec:background}

\subsection{Spectro-Computational Analysis}
Information theory establishes communication limits as functions of channel resources such as bandwidth and power \cite{shannon1948}. It does not directly account for the time required to execute the algorithms that realize a waveform. Consider a symbol carrying $B$ useful bits over a channel interval $T_{\mathrm{sym}}$. Its conventional PHY rate is
\begin{equation}
R=\frac{B}{T_{\mathrm{sym}}}.
\label{eq:conventional-rate}
\end{equation}
Immediately above the PHY, the first decoded bits are available only after transmit processing, channel transmission, and receive processing. SC throughput is therefore defined as \cite{queiroz2024,itu}
\begin{equation}
R_{\mathrm{SC}}=
\frac{B}{T_{\mathrm{comp,TX}}+T_{\mathrm{sym}}+T_{\mathrm{comp,RX}}},
\label{eq:sc-throughput}
\end{equation}
where $T_{\mathrm{comp,TX}}$ and $T_{\mathrm{comp,RX}}$ are the transmitter and receiver processing times, respectively. Dividing $R_{\mathrm{SC}}$ by bandwidth yields an SC spectral-efficiency metric.

Equation~\eqref{eq:sc-throughput} exposes computation-limited operation: increasing the nominal rate or bandwidth may require enough additional processing to offset the expected communication gain \cite{queiroz2024,itu}. It represents the delivery latency of one unit. A multi-unit opportunity additionally benefits from overlap among pipeline stages, which motivates the extension in Section~\ref{sec:model}.

\subsection{Software-Implemented Physical Layers}
Sora demonstrated real-time IEEE 802.11a/b/g processing on general-purpose multicore processors and reported detailed costs for individual PHY blocks \cite{sora}. Its optimizations include lookup tables, single-instruction multiple-data execution, dedicated cores, and pipelining. These data make Sora useful for comparing modulation and coding modes on a common platform.

More recent implementations illustrate the growing computational demands of programmable PHYs. A frequency-domain IEEE 802.11ac transceiver implementation on a very-long-instruction-word (VLIW) baseband processor with single-instruction multiple-data (SIMD) execution evaluated 256-ary quadrature amplitude modulation (256-QAM), low-density parity-check (LDPC) coding, and multiple-input multiple-output (MIMO) configurations \cite{aghababaeetafreshi2018}. MIMO-SoftiPHY implements a fifth-generation New Radio (5G-NR) uplink PHY integrated with OpenAirInterface and compatible with the O-RAN 7.2x functional split; it supports multiuser MIMO (MU-MIMO) and linear and nonlinear detection on general-purpose processors \cite{mimo-softiphy}. These studies establish the practical relevance of the computational constraint, although their reported measurements do not expose symmetric transmit and receive times for the same unit and platform. Section~\ref{sec:data-selection} explains why Sora is therefore used for the complete numerical application.

\section{Expanded Spectro-Computational Throughput Model}
\label{sec:model}
Let $D$ denote the useful-bit payload of a transmission opportunity and $B_m$ the number of useful bits carried by a fully occupied OFDM symbol in mode $m$. We refer to the consecutive symbols used for one opportunity as a PHY burst. Its required length is
\begin{equation}
S_m(D)=\left\lceil\frac{D}{B_m}\right\rceil.
\label{eq:number-symbols}
\end{equation}
When $D$ is not a multiple of $B_m$, the final symbol contains padding. Padding consumes processing and channel resources but is excluded from the useful-bit numerator.

The first symbol must traverse the three stages before its decoded bits become available above the PHY. Its end-to-end latency is
\begin{equation}
T_{\mathrm{first},m}=T_{\mathrm{comp,TX},m}+T_{\mathrm{sym},m}+T_{\mathrm{comp,RX},m}.
\label{eq:first-symbol}
\end{equation}
Once the stages operate concurrently, consecutive decoded symbols become available at the cadence imposed by the slowest stage:
\begin{equation}
T_{\mathrm{pipe},m}=\max\!\left\{T_{\mathrm{comp,TX},m},T_{\mathrm{sym},m},T_{\mathrm{comp,RX},m}\right\}.
\label{eq:pipeline-time}
\end{equation}

\begin{figure}[!t]
\centering
\includegraphics[width=0.96\textwidth]{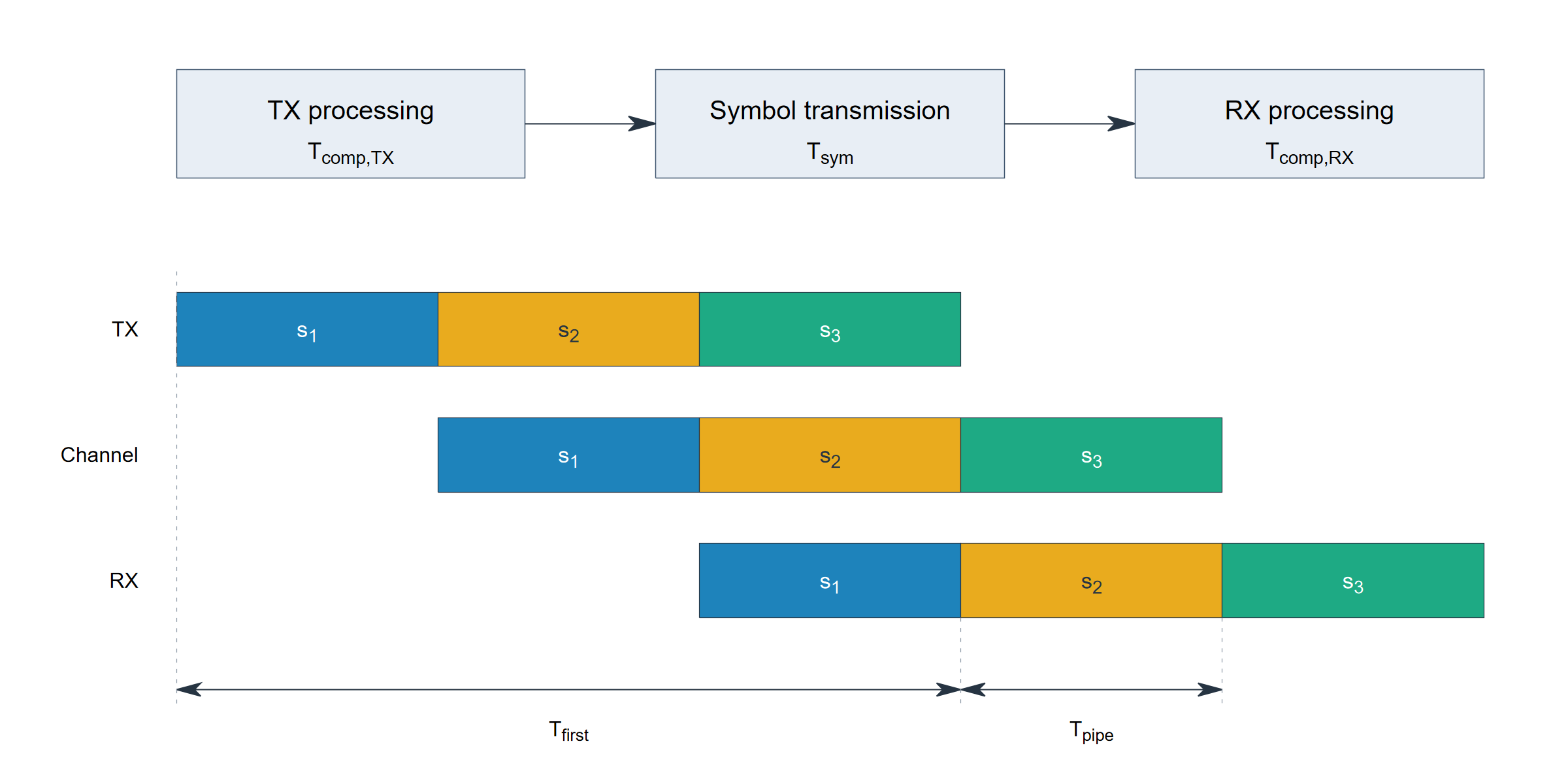}
\caption{Three-stage PHY pipeline for three consecutive symbols. Colors identify the symbols as they advance through TX processing, symbol transmission, and RX processing. The interval $T_{\mathrm{first}}$ extends from the start of TX processing for $s_1$ to the completion of its RX processing, whereas $T_{\mathrm{pipe}}$ is the interval between consecutive decoded outputs. Block widths are schematic and do not represent the relative stage durations. In general, $T_{\mathrm{pipe}}=\max\{T_{\mathrm{comp,TX}},T_{\mathrm{sym}},T_{\mathrm{comp,RX}}\}$. The mode index $m$ is omitted for clarity.}
\label{fig:pipeline}
\end{figure}

Accordingly, the end-to-end time required to deliver the complete PHY burst is
\begin{equation}
T_{\mathrm{total},m}(D)=T_{\mathrm{first},m}+
\left[S_m(D)-1\right]T_{\mathrm{pipe},m},
\label{eq:total-time}
\end{equation}

The expanded SC throughput is
\begin{equation}
R_{\mathrm{SC},m}^{\mathrm{exp}}(D)=
\frac{D}{T_{\mathrm{first},m}+\left[S_m(D)-1\right]T_{\mathrm{pipe},m}}.
\label{eq:expanded-throughput}
\end{equation}

For $S_m(D)=1$ and $D=B_m$, \eqref{eq:expanded-throughput} reduces to \eqref{eq:sc-throughput}. For fully occupied symbols and increasing $S_m$, the first-unit cost is amortized and throughput approaches $B_m/T_{\mathrm{pipe},m}$. Equation~\eqref{eq:expanded-throughput} therefore links short, computation-sensitive opportunities to the long-stream pipeline regime.

Consider modes $a$ and $b$ evaluated at the same $D$, with $R_{\mathrm{PHY},b}>R_{\mathrm{PHY},a}$. A computational-complexity anomaly occurs when
\begin{equation}
R_{\mathrm{SC},b}^{\mathrm{exp}}(D)<R_{\mathrm{SC},a}^{\mathrm{exp}}(D),
\label{eq:anomaly}
\end{equation}
or, equivalently, $T_{\mathrm{total},b}(D)>T_{\mathrm{total},a}(D)$. The equality of useful-bit payloads is essential; otherwise, an apparent inversion could be caused by comparing different delivered information quantities.

The model assumes approximately constant processing time per unit and sufficient independence to overlap consecutive units. It excludes preambles, medium-access contention, acknowledgments, retransmissions, and upper-layer headers. These terms can be added to \eqref{eq:total-time} for a specific protocol, but excluding them isolates the PHY computational effect examined here.

\section{Methodology and Data Selection}
\label{sec:method}

\subsection{Candidate Implementations}
\label{sec:data-selection}
Table~\ref{tab:implementations} summarizes the three software-implemented PHY studies considered. A direct application of \eqref{eq:expanded-throughput} requires, for the same processing unit, its useful-bit capacity, channel duration, and separable transmit and receive processing times.

\begin{table}[t]
\caption{Software-Implemented PHY Studies Considered in the Data-Selection Procedure}
\label{tab:implementations}
\centering
\small
\setlength{\tabcolsep}{5pt}
\renewcommand{\arraystretch}{1.12}
\begin{tabularx}{\textwidth}{
    >{\RaggedRight\arraybackslash}p{0.16\textwidth}
    >{\RaggedRight\arraybackslash}p{0.19\textwidth}
    >{\RaggedRight\arraybackslash}p{0.25\textwidth}
    >{\RaggedRight\arraybackslash}X}
\toprule
\textbf{Implementation} &
\textbf{Waveform and platform} &
\textbf{Reported computational evidence} &
\textbf{Use in this study}\\
\midrule
Sora/SoftWiFi \cite{sora} & IEEE 802.11a/b/g on a 2.66-GHz multicore general-purpose processor & Per-block costs for transmit and receive chains, a common platform across modes, and a 4-$\mu$s OFDM symbol & Selected for the numerical application because the required quantities can be reconstructed consistently. \\
\addlinespace[2pt]
IEEE 802.11ac PHY \cite{aghababaeetafreshi2018} & 80-MHz, 256-QAM, LDPC-coded MIMO on a VLIW/SIMD baseband processor & Detailed 
cycle counts for frequency-domain blocks; FFT/inverse FFT (IFFT) operations are assigned to dedicated hardware, and operating frequency changes across scenarios & Used to characterize computational scaling, but excluded from the end-to-end calculation because the complete software transmit/receive chains are unavailable. \\
\addlinespace[2pt]
MIMO-SoftiPHY \cite{mimo-softiphy} & 5G-NR MU-MIMO on an 18-core general-purpose processor, integrated with O-RAN split 7.2x & Uplink-chain latency and real-time feasibility per 0.5-ms slot & Used as an O-RAN-relevant implementation reference, but excluded from the symmetric calculation because separate transmit and receive times are not reported. \\
\bottomrule
\end{tabularx}
\end{table}

The IEEE 802.11ac implementation does not include the software FFT and inverse FFT in its measured chain \cite{aghababaeetafreshi2018}, even though transform complexity can dominate OFDM processing and scales approximately as $N\log_2N$ for $N$ subcarriers \cite{itu}. MIMO-SoftiPHY reports uplink processing by slot rather than matching TX and RX times for one unit \cite{mimo-softiphy}. Imposing missing values would introduce platform-dependent assumptions that could dominate the comparison. Sora is thus the only candidate used quantitatively.

\subsection{Sora Modes and Processing-Time Reconstruction}
The selected Sora modes are 24 Mbit/s with 16-QAM and code rate $1/2$, 48 Mbit/s with 64-QAM and code rate $2/3$, and 54 Mbit/s with 64-QAM and code rate $3/4$. Each mode uses 48 data subcarriers and a symbol duration $T_{\mathrm{sym}}=4~\mu$s \cite{sora}. Hence,
\begin{equation}
B_m=R_{\mathrm{PHY},m}T_{\mathrm{sym}},
\label{eq:bits-per-symbol}
\end{equation}
which yields 96, 192, and 216 useful bits per symbol, respectively.

Table~\ref{tab:costs} gives the optimized block costs extracted from Sora. The source reports scrambling/descrambling cost for the 54-Mbit/s mode. The same 40.29 million cycles/s value is conservatively used for the other modes. The 48-Mbit/s soft-demapping cost is set equal to the 54-Mbit/s value because both modes use 64-QAM and process the same number of coded bits per symbol.

\begin{table}[t]
\caption{Selected Sora Block Costs in Million Cycles/s \cite{sora}}
\label{tab:costs}
\centering
\small
\setlength{\tabcolsep}{3pt}
\begin{tabular}{c@{\hspace{5pt}}rrrrr}
\toprule
\textbf{Rate} & \textbf{Scr.} & \textbf{Enc.} & \textbf{FFT/} & \textbf{Soft} & \textbf{Vit.} \\
\textbf{(Mbit/s)} & \textbf{/descr.} & & \textbf{IFFT} & \textbf{demap.} & \\
\midrule
24 & 40.29 & 18.15 & 459.52 & 46.55 & 1408.93 \\
48 & 40.29 & 37.21 & 459.52 & 98.75 & 2422.04 \\
54 & 40.29 & 56.23 & 459.52 & 98.75 & 2573.85 \\
\bottomrule
\end{tabular}
\end{table}

Let $C_i$ be a block cost in cycles/s and $f_{\mathrm{CPU}}$ the central processing unit (CPU) frequency. Transmitter processing sums scrambling, convolutional encoding, and inverse FFT costs:
\begin{equation}
T_{\mathrm{comp,TX},m}=
\frac{C_{\mathrm{scr},m}+C_{\mathrm{enc},m}+C_{\mathrm{IFFT}}}{f_{\mathrm{CPU}}}
T_{\mathrm{sym}}.
\label{eq:tx-processing}
\end{equation}
Sora assigns the Viterbi decoder to one dedicated core and executes the other receiver blocks on a second core. The receive time is therefore the larger parallel branch:
\begin{equation}
T_{\mathrm{comp,RX},m}=\max\!\left\{
\begin{aligned}
&\frac{C_{\mathrm{vit},m}}{f_{\mathrm{CPU}}}T_{\mathrm{sym}},\\
&\frac{C_{\mathrm{descr},m}+C_{\mathrm{FFT}}+C_{\mathrm{demap},m}}
{f_{\mathrm{CPU}}}T_{\mathrm{sym}}
\end{aligned}
\right\}.
\label{eq:rx-processing}
\end{equation}
The calculation uses $f_{\mathrm{CPU}}=2.66$ GHz, matching the Sora evaluation platform \cite{sora}.

\section{Numerical Results and Anomaly Identification}
\label{sec:results}

\subsection{Processing Times}
Table~\ref{tab:times} reports the values obtained from \eqref{eq:tx-processing} and \eqref{eq:rx-processing}. All modes meet the 4-$\mu$s per-stage real-time constraint when the Viterbi decoder uses a dedicated core. The channel interval is therefore the slowest pipeline stage, and $T_{\mathrm{pipe},m}=4~\mu$s for all three modes.

\begin{table}[t]
\caption{Calculated Processing Times for the Selected IEEE 802.11a/g Modes}
\label{tab:times}
\centering
\small
\setlength{\tabcolsep}{3.5pt}
\begin{tabular}{crrr}
\toprule
\textbf{Rate} & $\boldsymbol{T_{\mathrm{comp,TX}}}$ & $\boldsymbol{T_{\mathrm{comp,RX}}}$ & $\boldsymbol{T_{\mathrm{first}}}$ \\
\textbf{(Mbit/s)} & \textbf{($\boldsymbol{\mu}$s)} & \textbf{($\boldsymbol{\mu}$s)} & \textbf{($\boldsymbol{\mu}$s)} \\
\midrule
24 & 0.779 & 2.119 & 6.898 \\
48 & 0.808 & 3.642 & 8.450 \\
54 & 0.836 & 3.870 & 8.707 \\
\bottomrule
\end{tabular}
\end{table}

If all receive blocks were executed sequentially on one core, the estimated receive times would be 2.940, 4.542, and 4.771 $\mu$s for 24, 48, and 54 Mbit/s, respectively. The two faster modes would exceed the symbol interval. Parallelization is therefore not merely an optimization in this implementation; it is required for real-time operation. Even after parallelization, receive time rises from 2.119 to 3.870 $\mu$s between the 24- and 54-Mbit/s modes, mainly because of Viterbi decoding.

\begin{figure}[t]
\centering
\includegraphics[width=0.98\columnwidth]{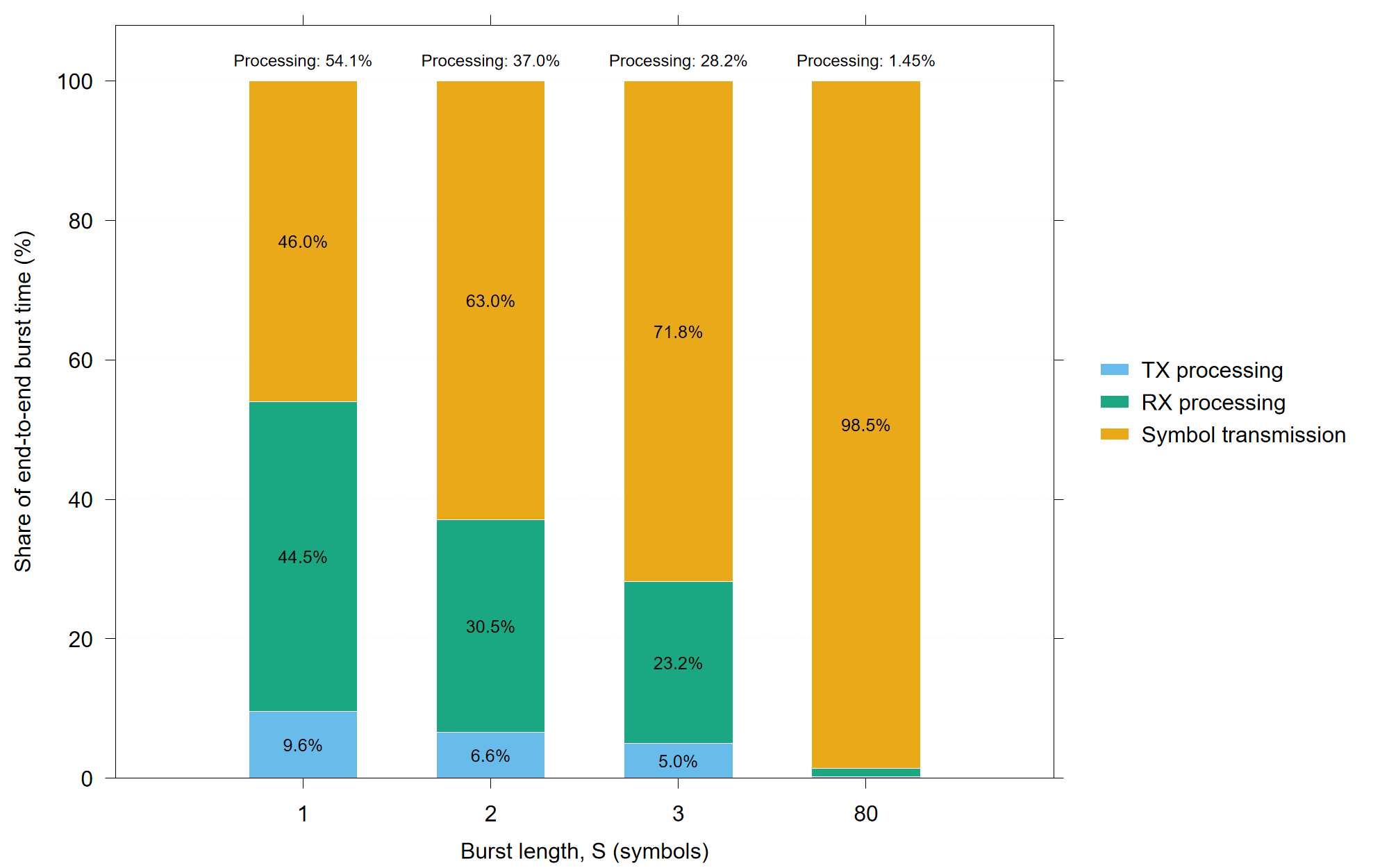}
\caption{Percentage decomposition of the end-to-end burst time for the 54-Mbit/s mode at selected burst lengths. Each 100\%-stacked bar separates TX processing, RX processing, and symbol transmission; labels above the bars give the combined processing share.}
\label{fig:processing-share}
\end{figure}

Because symbol transmission is the slowest stage for all three modes, the total time in this numerical case can be decomposed without double counting as
\begin{equation}
T_{\mathrm{total},m}(D)=T_{\mathrm{comp,TX},m}+S_m(D)T_{\mathrm{sym}}+T_{\mathrm{comp,RX},m}.
\label{eq:time-decomposition}
\end{equation}
Figure~\ref{fig:processing-share} reports the corresponding shares for the 54-Mbit/s mode, which has the largest processing latency. Processing accounts for 54.1\% of the total time when $S=1$, exceeding the contribution of symbol transmission. Its share falls to 37.0\% at $S=2$, 28.2\% at $S=3$, and 1.45\% at $S=80$. This transition directly illustrates how a short burst is computation-limited, whereas a long burst amortizes pipeline fill and drain latency.

\subsection{Expanded Throughput}
\begin{figure}[t]
\centering
\includegraphics[width=0.97\textwidth]{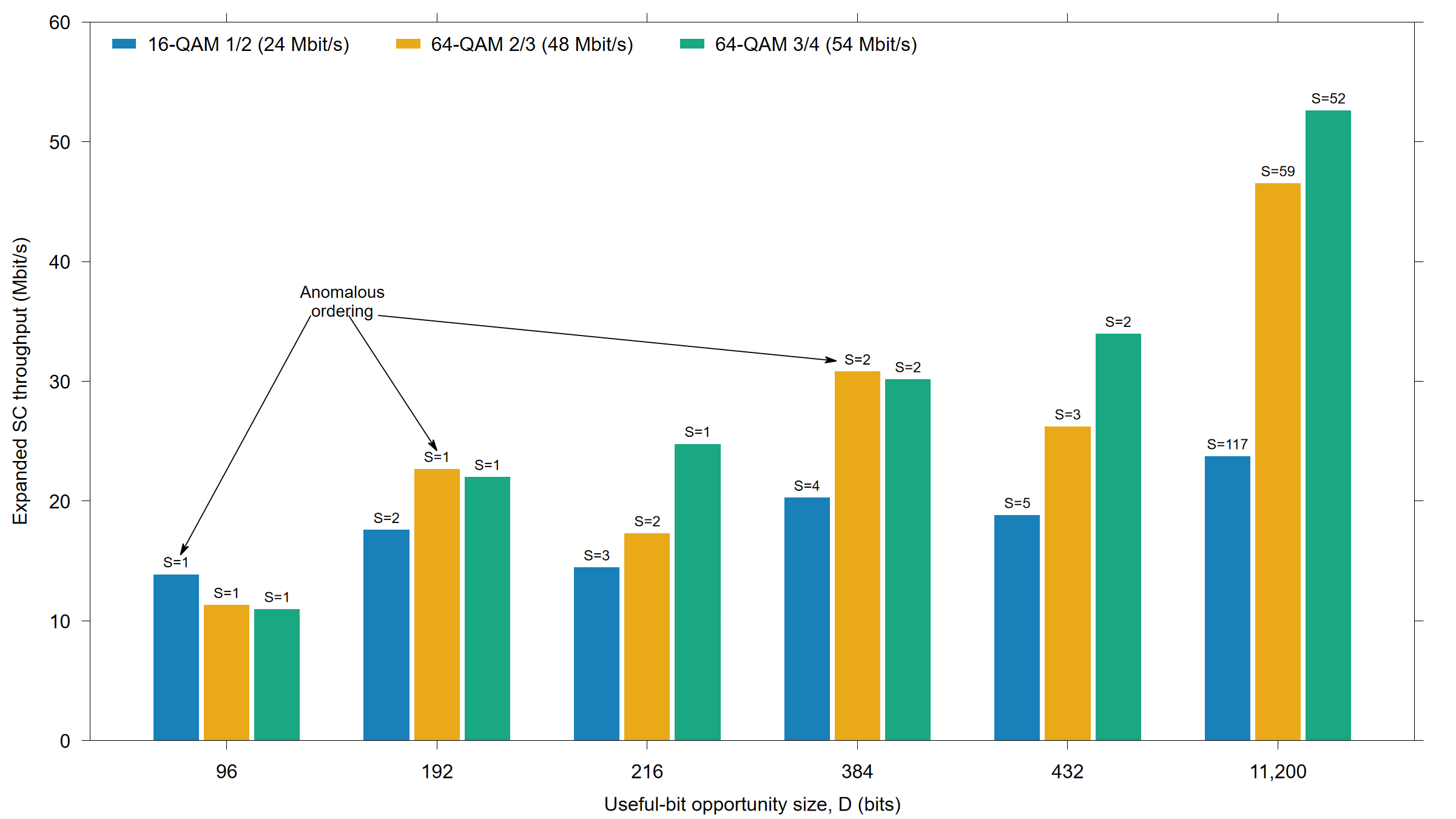}
\caption{Expanded SC throughput for selected useful-bit opportunity sizes. Labels above the bars give the number of symbols $S$ in the burst. Arrows identify cases in which a lower nominal PHY rate provides higher effective throughput.}
\label{fig:throughput-bars}
\end{figure}

Figure~\ref{fig:throughput-bars} evaluates \eqref{eq:expanded-throughput} for six opportunity sizes. The first five values expose symbol-count transition points, and the labels above the bars give the required burst length. The 11,200-bit opportunity represents 1,400 bytes without adding protocol headers or other overheads.

The selected opportunity sizes illustrate the effect of symbol-count transitions on expanded SC throughput. While $D$ remains within the capacity of the same number of symbols, throughput increases with the useful payload. Once that capacity is exceeded, one additional symbol is required and the total time increases by 4 $\mu$s. The resulting ranking therefore depends on the alignment between opportunity size and symbol capacity, not only on nominal rate.

At $D=96$ bits, all modes use one symbol. The 48- and 54-Mbit/s modes leave more capacity unused and incur higher computational latency, so the 24-Mbit/s mode is fastest. An anomalous ordering also occurs between the two 64-QAM modes: the 48-Mbit/s mode reaches 11.361 Mbit/s, slightly exceeding the 11.026 Mbit/s of the 54-Mbit/s mode. At $D=192$ bits, both 48 and 54 Mbit/s require one symbol, but the 48-Mbit/s mode provides 22.723 Mbit/s and exceeds the 22.052 Mbit/s of the nominally faster mode. At $D=216$ bits, the 54-Mbit/s mode fits the payload into one symbol, whereas the 48-Mbit/s mode needs two; the symbol saving reverses the result. The anomalous ordering observed at 192 bits reappears at 384 bits, whereas the symbol-count advantage observed for the 54-Mbit/s mode at 216 bits reappears at 432 bits.

An exhaustive evaluation of $D$ from 4 to 2,112 bits in 4-bit increments confirms that the representative inversions in Figure~\ref{fig:throughput-bars} are not isolated. The 48-Mbit/s mode outperforms the 54-Mbit/s mode over 4--192, 220--384, 436--576, 652--768, 868--960, 1,084--1,152, 1,300--1,344, and 1,516--1,536 bits. In each interval, both modes use the same number of symbols, but the first-symbol latency of the 54-Mbit/s mode is 0.257 $\mu$s greater.

At 1,536 bits, both modes still require eight symbols. Their total times are 36.450 and 36.707 $\mu$s, and their throughputs are approximately 42.140 and 41.845 Mbit/s for the 48- and 54-Mbit/s modes, respectively. At 1,540 bits, the 48-Mbit/s mode requires a ninth symbol while the 54-Mbit/s mode remains at eight, restoring the expected ordering. Beyond 1,536 bits, the two modes do not again require the same symbol count in the evaluated discrete sequence, and the anomaly does not reappear.

For 11,200 bits, the initial latency is amortized over dozens of symbols. Throughput approaches the nominal ordering: 23.784, 46.579, and 52.655 Mbit/s for the 24-, 48-, and 54-Mbit/s modes. The expanded formulation is therefore most consequential for short opportunities in which symbol granularity and processing time are not amortized.

\section{Discussion and Limitations}
\label{sec:discussion}
The anomaly has a direct link-adaptation implication. A controller that selects the highest feasible nominal modulation and coding rate can reduce useful throughput when two candidate modes require the same number of symbols. In such intervals, the extra computational budget of the more aggressive mode buys no reduction in channel occupancy. That budget could instead support stronger channel coding, more reliable detection, or additional decoder capacity. When the higher rate removes an entire symbol, however, the granularity gain can dominate the added computation and justify the more aggressive mode.

This result is relevant to O-RAN even though the numerical case study predates modern O-RAN deployments. O-RAN-compatible PHYs execute increasingly complex signal-processing chains on shared programmable infrastructure \cite{o-ran,mimo-softiphy}. Their scheduling and adaptation policies can observe processor load and measured latency in addition to channel state. Equation~\eqref{eq:expanded-throughput} supplies a compact objective for such cross-layer decisions, provided that platform-specific transmit and receive timings are measured.

The numerical values should be interpreted as throughput for the modeled PHY processing unit, not as complete application throughput. Preambles, interframe spaces, medium access, acknowledgments, headers, and retransmissions are excluded. The small $D$ values are analytical opportunities used to expose symbol granularity; they are not claimed to be complete Wi-Fi frames. In addition, Sora reports selected PHY block costs rather than every operation in a modern transceiver. The scrambling and 48-Mbit/s soft-demapping assumptions described in Section~\ref{sec:method} also introduce estimation uncertainty.

These limitations do not change the internal comparison: all modes use the same platform, symbol duration, processing architecture, and block set. They do limit external generalization. A modern validation should measure complete transmit and receive paths on a 5G-NR O-RAN platform, report confidence intervals over repeated timing samples, and include processor contention, memory-transfer effects, energy, and reliability. The model can then be extended with protocol overheads and used in an adaptation mechanism that jointly selects modulation, coding, symbol count, and compute allocation.

\section{Conclusion}
\label{sec:conclusion}

In this work, we characterized a throughput anomaly that can affect software-based 
PHY implementations, which play a key enabling role in Open Radio Access Network (O-RAN) systems. 
The anomaly is most pronounced when a single PHY transmission unit (i.e., a symbol) suffices 
to convey the entire offered load--a condition that becomes increasingly likely as nominal PHY bit rates increase. 
In this limiting case, intersymbol pipelining across the modulation, transmission, and demodulation stages 
cannot be exploited, and the associated processing latency accounts for a substantial fraction of the total delivery time. 
This effect can be further amplified because higher-rate signal configurations often increase the relevant signal 
dimensions and, consequently, the computational cost of PHY algorithms, some of which scale superlinearly with these dimensions. 
Once this processing latency is properly accounted for, a nominally higher-rate configuration may deliver lower 
effective PHY throughput than a nominally slower configuration, contrary to the performance ranking implied by conventional bit-rate metrics.

We demonstrated that this anomaly constitutes a practical manifestation of the computation-limited regime 
formalized by the SC framework introduced in~\cite{queiroz2024}. In this regime, asymptotic communication 
performance is ultimately constrained by computational resources, even when conventional channel resources, 
such as bandwidth and transmit power, become arbitrarily abundant. To characterize this behavior for bursts of 
different lengths, we extended the SC framework by explicitly parameterizing the number of symbols per transmission 
opportunity. Our findings reveal a practical concern for emerging very-high-throughput software-based PHY implementations, 
which are increasingly likely to operate with short bursts comprising only a few high-density symbols. As future work,
 we plan to develop link-adaptation algorithms that explicitly account for symbol-processing latency when selecting the 
signal configuration that maximizes effective throughput. 

\section*{Acknowledgment}
This work has been partially funded by the project Advanced Multimodal Sensing
(AIMS), supported by the Advanced Knowledge Center in Immersive Technologies
(AKCIT), with financial resources from the PPI IoT of the MCTI, under Grant
057/2023, signed with EMBRAPII. The authors are also grateful to the
Funda\c{c}\~ao de Amparo \`a Pesquisa do Estado de Goi\'as (FAPEG) for the
financial support provided for this research under Grant 64448878/2024.

The authors used OpenAI ChatGPT, solely for grammatical review
of the manuscript.

\bibliographystyle{unsrt}
\bibliography{references}

\section*{Author Biographies}

\subsection*{Gabriel Alessi Posonski}
is currently pursuing
the B.S. degree in computer science at the Federal University of
Technology--Paran\'a (UTFPR), Ponta Grossa, Brazil. He also works as a
software developer. His research interests include software-defined radio,
open radio access networks, wireless physical-layer signal processing, and
spectro-computational analysis.

\subsection*{Saulo Queiroz}
is an Associate Professor with the Department of Computer Science, Federal
University of Technology--Paran\'a (UTFPR), Brazil. He received the Ph.D.
degree with distinction and honors from the University of Coimbra, Portugal.
During his academic training, he contributed to open-source networking
projects, including Google Summer of Code initiatives. Over the last decade,
he has taught computer science subjects including design and analysis of
algorithms, data structures, and communication signal processing. His current
research interests include networking and signal processing for wireless
communications.

\subsection*{Jo\~ao P. Vilela}
is an Associate Professor with the Department of Computer Science, University
of Porto, Portugal, and center coordinator at INESC TEC. He received the Ph.D.
degree in computer science from the University of Porto in 2011. He was a
Professor with the University of Coimbra and a Visiting Researcher with Georgia
Tech and MIT, USA. He has coordinated and participated in several national,
bilateral, and European-funded projects in security and privacy. His main
research interests are in security and privacy of networked and intelligent
systems, with applications in next-generation wireless and mobile environments.
Key topics include security and privacy of 6G networks and integrated sensing
and communication, as well as trustworthy machine learning---including
privacy-preserving federated learning and generative models---and automated,
scalable privacy-enhancing technologies.

\subsection*{Benjamin Koon Kei Ng}
received the B.A.Sc., M.A.Sc., and Ph.D. degrees in engineering science and
electrical engineering from the University of Toronto, Toronto, ON, Canada,
in 1996, 1998, and 2002, respectively. From 2005 to 2009, he was a Senior
Communications Engineer with Radiospire Networks Inc., Boston, MA, USA,
focusing on ultrawideband and millimeter-wave technologies. He joined Macao
Polytechnic University, Macao, China, in 2010, where he is currently an
Associate Professor with the Faculty of Applied Sciences. His research
interests include wireless communications and signal processing, with an
emphasis on MIMO, NOMA, and machine-learning technologies.

\subsection*{Chan-Tong Lam}
received the B.Sc. (Eng.) and M.Sc. (Eng.) degrees from Queen's University,
Kingston, ON, Canada, in 1998 and 2000, respectively, and the Ph.D. degree
from Carleton University, Ottawa, ON, Canada, in 2007. He is currently an
Associate Professor with the Faculty of Applied Sciences, Macao Polytechnic
University, Macao, China. From 2004 to 2007, he participated in the European
Wireless World Initiative New Radio (WINNER) Project. His research interests
include mobile wireless communications, machine learning in communications,
and computer vision for smart cities.

\subsection*{Edmundo Monteiro}
is a Full Professor with the University of Coimbra, Portugal. He has more than
30 years of research experience in computer communications, wireless
networks, quality of service and experience, network and service management,
and computer and network security. He has participated in many Portuguese,
European, and international research projects and initiatives. His publication
record includes more than 200 publications in journals, books, and
international refereed conferences, and he has co-authored nine international
patents. He is a member of the Editorial Board of \emph{Wireless Networks}
(Springer), a Senior Member of the IEEE Communications Society, a member of
the ACM Special Interest Group on Communications, and the Portuguese
representative in IFIP TC6 (Communication Systems).

\end{document}